\documentclass[runningheads]{llncs}
\usepackage{subcaption}
\usepackage{graphicx}
\usepackage{wrapfig}
\begin{document}

\title{Making Sense of Metadata Mess: Alignment \& Risk Assessment for Diatom Data Use Case}
\author{Kio Polson\inst{1}\orcidID{0009-0002-3087-0739} \and
Marina Potapova\inst{2}\orcidID{0000-0001-5143-5213} \and
Uttam Meena\inst{3} \and
Chad Peiper\inst{1}\orcidID{0009-0004-9911-0951} \and
Joshua Brown\inst{4}\orcidID{0000-0003-1227-6429} \and
Joshua Agar\inst{3}\orcidID{0000-0001-5411-4693} \and
Jane Greenberg \inst{1}\orcidID{0000-0001-7819-5360}
}
\institute{Metadata Research Center, Drexel University, Philadelphia, PA, USA \\
\email{\{kp3272,cep98,jg3243\}@drexel.edu} \and
Academy of Natural Sciences of Drexel University, Philadelphia, PA, USA \\
\email{mp895@drexel.edu} \and
College of Engineering, Drexel University, Philadelphia, PA, USA \\
\email{um44@dragons.drexel.edu,jca92.drexel.edu} \and
Oak Ridge National Lab, Oak Ridge, TN, USA
\email{brownjs@ornl.gov}}

%\author{Jane Greenberg\inst{1} \and
%Scott McClellan\inst{1} \and
%Xintong Zhao\inst{1} \and
%Elijah J Kellner\inst{2} \and
%David Venator\inst{3}\and
%Haoran Zhao\inst{1}\and
%Jiacheng Shen\inst{1}\and
%Xiaohua Hu\inst{1}\and
%Yuan An\inst{1}}

%\institute
%{
%  \inst{}%
%  Metadata Research Center\\
%  Drexel University
%%  \and
%  \inst{}%
%  College of Science and Engineering\\
%  Winona State University
%  \and
%  \inst{}
%  McCormick School of Engineering\\
%  Northwestern University
%}

%
%\authorrunning{Greenberg et al.}
\maketitle              % typeset the header of the contribution
\begin{abstract}
Biologists study Diatoms, a fundamental algae, to assess the health of aquatic systems. Diatom specimens have traditionally been preserved on analog slides, where a single slide can contain thousands of these microscopic organisms. Digitization of these collections presents both metadata challenges and opportunities. This paper reports on metadata research aimed at providing access to a digital portion of the Academy of Natural Sciences' Diatom Herbarium, Drexel University. We report results of a 3-part study covering 1) a review of relevant metadata standards and a microscopy metadata framework shared by Hammer et al., 2) a baseline metadata alignment mapping current diatom metadata properties to standard metadata types, and 3) a metadata risk analysis associated with the course of standard data curation practices. This research is part of an effort involving the transfer of these digital slides to an new system, DataFed, to support global accessible. The final section of this paper includes a conclusion and discusses next steps.

\end{abstract}

\section{Introduction}

Creating and managing high quality metadata is essential for supporting digital life-cycle management and the FAIR principles \cite{Wilkinson2016}. These aims inform a current initiative involving the Academy of Natural Sciences’ Diatom Herbarium (ANS Diatom Herbarium), Drexel University. Diatoms are a fundamental algae, which remove carbon dioxide from the atmosphere through photosynthesis. Researchers focused on diatoms collect and preserve water samples on slides. They use microscopy technology to observe diatom microorganisms. Today, the ANS Diatom Herbarium contains well over 300,000 slides and is recognized as one of the most extensive diatom collections in the United States. In 2009, with the support of the U.S. National Science Foundation, the creators of this collection undertook a project to digitize 6,000 slides, a subset of the collection. The aim was to make the collection more globally accessible.

Indeed, the digitization of these diatom slides was an important step toward increased collection access and data sharing, although this activity revealed a number of significant challenges related to collection organization and curation. The main challenges stem from the fact that a diatom specimen is microscopic. Biological database software generally assumes that a slide simply captures a single specimen or organism, while a diatom slide may contain thousands of microscopic specimens. Given this challenge and the absence of a unified technology supporting project needs, the ANS team was required to develop a system with a unique combination of microscope hardware, digitization hardware, and digitization software. The workflows, metadata gathering, and processing activities were conducted in house, as software libraries accounting for this specific use case are not available.

In early 2024, a new effort was launched to enhance access to the ANS Diatom Herbarium and address these noted challenges. This current initiative involves information and computer scientists and diatom experts, with key aims to: 1) advance the management of the ANS Diatom Herbarium, and 2) support global access to this unique collection. The collaboration is connected with a more recent NSF project, ‘Development of a Platform for Accessible Data-Intensive Science and Engineering.’ A major component of the larger NSF project is to develop a metadata infrastructure for porting the digital collection to a developing system, called DataFed \cite{Stansberry2020}. This paper discusses the current metadata research activities specific to the ANS Diatome Herbarium collection. We report on our review of metadata standards, a baseline metadata alignment, and a metadata risk analysis. The sections that follow cover these three activities, followed by a discussion and a conclusion.

\section{Review of Metadata Standards}

Over the last 15 years, there’s been a significant increase in  metadata activities across science. One key motivator is to improve support for research reproducibility \cite{Baker2016}. As a result, a number of community-driven metadata standards, including efforts in microscopy have emerged. These relatively new standards typically gather a series of microscopy activities under a single unit known as an “imaging experiment.” Hammer et al.\cite{Hammer2021} explain that the evolving metadata landscape for an imaging experiment has the following three categories:
\begin{itemize}
    \item Experimental and Sample Metadata (documents sample preparation)
    \item Microscopy Metadata (documents image data acquisition)
    \item Analysis Metadata (documents image analysis)
\end{itemize}

These categories provide a framework that informs our review, whereby a digitized diatom slide, or a set of slides, fall specifically under the “imaging experiment” label. A chief goal here is to provide as much context as possible for the digitized slides, thus, enabling diatom researchers around the world to conduct analytical experiments using these published slides.

\section{Relevant Metadata Standards}

In recent years, there have been two major schemas relevant to microscopy: the Open Microscopy Environment (OME) \cite{OMEWebsite} and Digital Imaging and Communications in Medicine (DICOM) \cite{DICOMWebsite}. Darwin Core (DwC) \cite{Wieczorek2012}, which is used for scientific samples is also relevant for sample metadata.

\begin{figure}
\centering
\includegraphics[scale=0.65]{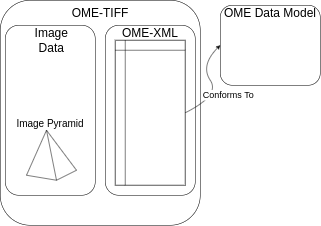}
\caption{High-level Overview of current OME Standards}\label{fig:ome}
\end{figure}

The OME appears is the most aligned with the needs of the diatom team. Currently, the main file format that the OME supports is called OME-TIFF \cite{OME-TIFF}. This file format stores image information and pixels in a TIFF file, bundled together with metadata that strictly adheres to the OME Data model \cite{OME-XML} in an XML serialization. Figure~\ref{fig:ome} presents a high-level overview of these OME standards.

While OME-TIFF is the current standard, recent efforts point to a new file format, OME-Zarr \cite{Moore2023}. This development is supported by working group focusing on OME Next Generation File Formats (OME-NGFF) \cite{NGFFWebsite}. Research indicates that OME-NGFF will have three major improvements when compared with OME-TIFF. First, instead of relying on TIFF based images that need to keep track of multidimensional images in one file, the new standard will move to a Zarr \cite{ZarrWebsite} based file format. Zarr is a file format where any number of imaging dimensions and image sizes can be handled using well-structured folders. This Zarr file format is also easier to store and access from cloud computing solutions. Second, instead of using XML based metadata, the OME Data Model aims to adhere to a Resource Description Framework (RDF) \cite{Kobayashi2019}. Third, the OME Data Model would be expanded to account for the changes in microscopy technology that has emerged since 2016. The most likely candidate for how the OME Data Model would be extended comes from a joint collaboration between the 4D Nucleome Initiative (4DN), the BioImaging North America (BINA) and the OME, to form the 4DN-BINA-OME (NBO) framework \cite{Hammer2021}.

In addition to these OME-TIFF updates, this effort emphasizes the need for extensibility in microscopy metadata. Underlying reasons include supporting quality and reproducibility \cite{Nelson2021}, complex biomedical tissue imaging \cite{Schapiro2022}, and basic information for interoperable archiving \cite{Sarkans2021}.

In addition to OME, advances have also been made with DICOM as a metadata standard for microscopy \cite{Gupta2022}. DICOM has been used in the biomedical community and supports robust file format and file transfer protocols. OME lacks the same level of rigor compared to DICOM, especially regarding the transfer of data at time of publication. Overall, DICOM offers a more developed suite of standards, although it does not fully support the needs of the ANS Diatom Herbarium.

Finally, it is important to mention the Darwin Core. The Darwin Core (DwC) is globally recognized and used for scientific samples, including museum specimens, and serves as the baseline standard for the ANS Diatom Herbarium. The diatom team has extended this standard in-house to improve the metadata quality of their collection \cite{Potapova2022}.

\section{Baseline Metadata Alignment}

The baseline metadata alignment that follows builds on the Hammer et al. framework. Hammer et al. present a set of subdivisions (also known as types) of metadata beyond \textit{sample metadata}, \textit{microscopy metadata}, and \textit{analysis metadata}. They further break down "microscopy metadata" into two subcategories, \textit{provenance metadata} and \textit{quality control metadata}. "Provenance metadata" in microscopy is further subdivided into three more categories of metadata: \textit{microscope hardware specifications metadata}, \textit{image acquisition settings metadata}, and \textit{image structure metadata}. The ANS Diatom Herbarium work has also identified and captured “scanning metadata,” which is not part of the Hammer et al. framework. Below is a tree structure of  Hammer et al. framework and how it aligns with our diatom use case, followed by a key Table~\ref{tab:align_key} to support interpretation:

\begin{itemize}
    \item \textbf{Sample Metadata (D)}
    \item Microscopy Metadata
    \begin{itemize}
        \item Provenance Metadata
        \begin{itemize}
            \item \textbf{Microscope Hardware Specifications Metadata (D)}
            \item \textbf{Image Acquisition Settings Metadata (D)}
            \item \textbf{Image Structure Metadata (D)}
            \item \textit{Scanning Metadata (X)}
        \end{itemize}
        \item Quality Control Metadata
    \end{itemize}
    \item Analysis Metadata
\end{itemize}

\begin{table}
    \centering
    \caption{Key of Metadata Alignment Tree}
    \label{tab:align_key}
    \begin{tabular}{|l|l|l|}
        \hline
        Style & Part of Hammer et al. Framework & Part of Diatom Metadata \\
        \hline
        $\cdot$ Normal Text & Yes & No \\
        $\cdot$ \textbf{Bold (D)} & Yes & Yes \\
        $\cdot$ \textit{Italics (X)} & No & Yes \\
        \hline
    \end{tabular}
\end{table}

\subsection{Sample Metadata}

Sample metadata includes any descriptive, structural, or provenance metadata that relates to information about the water on glass slides used for viewing diatoms under a microscope. Sample metadata has been stored in a database that capture's the complexity of this information. Attributes like: water sample original location, time and date of collection, name of person who collected sample, ID given to (and printed on) slide, etc.

\begin{figure}
\includegraphics[width=\textwidth]{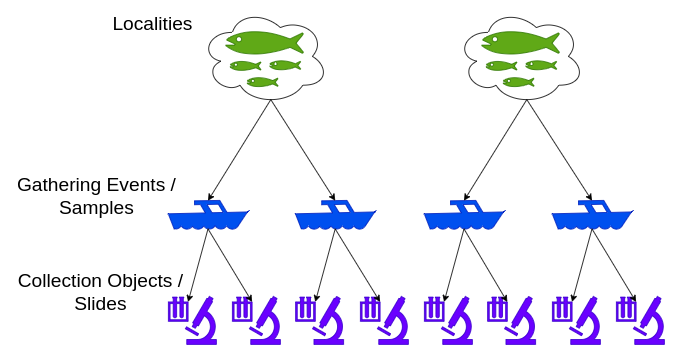}
\caption{Sample Metadata Data Structure for the ANS Diatom Herbarium} \label{fig:tax}
\end{figure}

The information contained in this database is extremely rich and critically important to the diatom team. The structure of this metadata is loosely based on the Darwin Core metadata schema, with several modifications to account for the microscopic nature of a specimen. At a high-level view, the most general entity is the ‘locality’ which defines which body of water is being discussed. Each ‘locality’ can have multiple ‘gathering events.’ Each ‘gathering event’ corresponds to a sample of water. Two ‘gathering events’ can occur at the same 'locality' and at the same time, at different body of water locations, or at the same exact location at a 'locality' but collected at different times. Finally, each sample from a gathering event can produce multiple ‘collection objects.’ A collection object can be a slide prepared for a microscope or a sub-sample where the water has gone through some ‘preparation’ and placed in a smaller container. Each ‘collection object’ has an ID printed directly on the physical object, and this ID is the same one that is in the database. Furthermore, this same ID is also associated with the scan of the slide, which is discuss later. The basic taxonomic structure described above is presented in Figure~\ref{fig:tax}.

Sample metadata is contained in a MySQL Database. The database has at most twenty-nine tables, although the key architect is not accessible. Unfortunately, the lack of contact means it is unclear how many of these tables are used by the schema, or are relics of previous versions that were never deleted. 

Additional challenges stem from File Maker Pro, which provides a front-end allowing users to view database contents. File Maker Pro has had a number of significant updates  since the database was first constructed and modified support for certain tasks. Given various project constraints, neither the database structure nor the templates generated in the previous File Maker Pro version have been updated for the current database. This situation presents backwards compatibility issues for ANS Diatom Herbarium team members seeking navigating their previously collected sample metadata. Templates supporting certain functionalities in File Maker Pro versions no longer offer the same support. As a result, diatom team members need to learn and develop new workarounds to accomplish their goals. One key challenge is batch editing sample metadata areas that need to be cleaned and updated.

\subsection{Microscope Hardware Specifications Metadata}

‘Microscope hardware specifications metadata’ contains descriptive metadata about the microscope hardware specification the diatom team used when they conducted the scan of a slide. The microscope hardware specification is critical to scientists studying slide scans, given the impact on interpreting and understanding diatom characteristics. The specification settings are not automatically and adequately captured by the scanning software. The scanning software records some parts of the microscope, like the focus dial. However, a human has to manually change the illumination of the stage and the lens magnification, and this specification metadata has to be recorded separately.

The diatom team has chosen an innovative way to ensure the physical settings and calibration of the microscope stay with the scan of a slide. Once a scan is completed, the diatom team needs to name the folder which contains all the files and images associated with a single scan. The name selected for this folder is one that combines these microscope hardware specifications metadata together, with an underscore '\_' between metadata elements. The precise order and inclusion of metadata changed over time, although the most recent version follows the following pattern.

\begin{verbatim}
    [Slide ID]_[Sample ID]_[Magnification]_[Illumination]_
    [Focal plane capture mode]_[Number of focal planes]_
    [Distance between focal planes]
\end{verbatim}

Following is an outline of these components and a brief definition:

\begin{description}
    \item[\texttt{Slide ID}] -- A unique series of letters and numbers printed on the physical slide and recorded in the Sample Metadata Database
    \item[\texttt{Sample ID}] -- A unique series of letters and numbers printed on the bottle of water which contains the water that had originally produced the slide in question
    \item[\texttt{Magnification}] -- Lens used for scan which magnifies the image
    \item[\texttt{Illumination}] -- The method used to illuminate the slide for imaging
    \item[\texttt{Focal plane capture mode}] -- Either ‘Zstack’ or ‘Zbest’. Defines strategy for capturing focal planes
    \item[\texttt{Number of focal planes}] -- When capturing multiple focal planes, indicates how many focal planes were captured
    \item[\texttt{Distance between focal planes}] -- Defines how gradual focal plane would change between images taken
\end{description}

Each scan compiles this information into the folder name. The exact schema changed over time, however, given the reliance on manual metadata generation and the and the titles were prone to human error. Still, this was a relatively effective way to keep some hard coded metadata to stay with each scan.

\subsection{Image Acquisition Settings Metadata}

For the diatom use case, digitization is a type of image acquisition. When digitizing diatom slides, there are additional software settings that the diatom experts can specify to ensure a clear image is captured. We  refer to these settings as the ‘image acquisition settings.’ The type of metadata includes a combination of technical, preservation and descriptive metadata  relating to the software used when creating a scan of a slide. The software produces technical metadata which details some descriptive and preservation metadata related to calibrating the software itself for a scan.

This calibration has many components that are recorded for each scan. Specifically, there are 107 different fields across six categories. For the Diatom team, once the software was calibrated for one slide, this calibration would usually work for following slides as well. Occasionally, something would change, though, and the software would need to be recalibrated in order to produce a professional scan. The majority of data captured is contained in an automatically generated file after a scan is complete. This file has relevant metadata that appears to be most at risk of getting lost in the process of data curation.

\subsection{Image Structure Metadata}

As the digitization of a slide is taking place, multiple images are captured that will later be stitched together. In the diatom use case, the component images are referred to as ‘tiles.’ These tiles can eventually be recompiled similar to a large jigsaw puzzle, usually twenty tiles by twenty tiles. In addition, in order to get the most clarity to see each diatom in a slide, it was necessary to take images on multiple focal planes. Thus, digitization not only captures tile images that would make up a single larger image, or one jigsaw puzzle, it has to recreate an additional dimension. The tiles frequently make up multiple full images or multiple jigsaw puzzles stacked one on top of the other. There could be a variable number of these image stacks according to the assessment of diatom experts to optimize how many diatoms are visible on each slide. All of this complexity needed to be recorded to document the image structure. Thus, the image structure metadata was any metadata that kept track of which tile needed to go where in the final stack of images.

This metadata was automatically generated by the software into a text file. This file captures the full complexity of where exactly each component image for a slide scan was taken in space and how that measurement in micrometers translates to the number of pixels in the slide. The file proved to be a critical piece of metadata for each scan when reconstructing the original image, and was also at risk of getting lost or deleted in the process of data curation.

\subsection{Scanning Metadata}

Finally, the scanning metadata is any descriptive or provenance metadata relating to the scanning and digitization event. Once the slide is on the stage, the microscope hardware is properly calibrated, and the digitization software is properly calibrated, a person can initiate the scanning process. This step of actually taking the scan also produces metadata.

An Excel spreadsheet was used for capturing this scanning metadata. For the diatom team accomplishing the scans of these slides, it is useful for them to have a method to record when a scan has been completed and for which slide. At first, this spreadsheet was very sparse and only had a boolean record of whether a slide had been scanned or not. Since our collaboration started, a new spreadsheet was created which also included fields of what date was a scan taken, who was the student who took the scan of a slide, along with some image acquisition settings metadata for redundancy. The image acquisition settings metadata was repeated in the Excel spreadsheet so that the folder name could be automatically generated to try to avoid some human error in the naming conventions. This amount of redundancy seemed appropriate for this use case.

\section{Metadata Risk Assessment}

The ongoing diatom collaboration made evident that the overall size and scope of this project introduces many technical issues that present potential road blocks to other research that seeks to use the ANS Diatom Herbarium digital slides. These include: 1. issues of storage space, 2. metadata loss, 3. metadata inconsistency, 4. suppression of metadata maintenance, and 5. missing metadata. Clear identification and articulation of these challenges in integral to addressing them and mitigating risk in future projects, given the speed at which technology changes.

\subsection{Storage Space and Potential Data Loss}

A single scan for microscopy imaging and digitization project is quite extensive. For example, the scan of a single slide regularly in the ANS Diatom Herbarium took up least twenty gigabytes of space, and this was a project, attempting to scan and digitize only 6,000 slides. This implies needing at least 100 terabytes of storage space somewhere. Unfortunately, the diatom team ran out of server space part way through the process of digitizing and had to start storing slides on external hard drives. The use external drives reduced the capacity for metadata backups and increases the opportunity for file corruption. This limitation required the team to find ways for each scan to take up less space, or to add more storage space.

Because some scans of slides are only stored on external hard drives, both data and metadata alike are at risk of being corrupted without being backed up. Whenever possible, original data and metadata should both be kept on working computers and backed up in multiple places to help with the preservation and sustainability of data.

\subsection{Metadata Loss from Image Conversion}

Each microscope imaging solution has its own proprietary file format for saving a microscope image. This means that for future use, the images have to be converted to a format which is easier to view. In microscopy, the popular solution is to use OME-TIFF. This is an open source file format specifically for making it easier to view microscopy images. However, there is no, one size fits all, way to take raw image files from unique microscope setups and easily turn them into OME-TIFF files. In the diatom case, lots of technical work has to be done in order to properly convert the images. This meant extra time and money spent on creating a workflow for this conversion.

The original solution that the diatom team had devised to convert scanned slides to OME-TIFF lost all image acquisition settings metadata and all the image structure metadata in the process. While this metadata might not have been a priority for the diatom team, it could still be important down the road for digital reconstruction or for the details that are available about the focal length. For this reason, it has proven important to consult with information scientists, or to have a metadata review process to ensure that no metadata gets lost in the conversion process.

\subsection{Metadata Inconsistency from Human Error}

As was mentioned earlier, the naming convention of the folders changed over time as the diatom team was honing in on their scanning methodology. These were not significant changes, but it does mean that the naming convention listed in this paper could not be relied upon for all scans of slides. This can interfere with future search functionality if left unchanged. With this deeper understanding of the metadata involved, it should then be possible to clean these folder names as a subproject of this initiative. Furthermore, creating systems and workflows that can minimize human error can also be important in the stewarding of metadata.

\subsection{Suppressed Metadata Maintenance from Bad User Interface (UI)}

As was mentioned earlier, the original architect of the MySQL Database containing the sample metadata was no longer in contact with the diatom team. For a considerable amount of time, this meant that the entire database was essentially inaccessible without the right credentials. Even once access was obtained, access was still limited by the changes to File Maker Pro that had occurred since the original File Maker Pro files had been created. These limitations made accessing, updating, and correcting the database incredibly difficult.

The diatom team includes committed stewards wit metadata expertise, although the User Interface (UI) tools have been cumbersome and make metadata update and maintenance tasks extremely challenges. While a new UI for the existing MySQL database is needed, development and implementation require resources and time, presenting a new set of challenges. Overall, this challenge underscores importance of good UI for proper metadata management.

\subsection{Missing Metadata}

Finally, the OME comparison work allowed us to identify metadata that was not currently being recorded. The OME data model allows for there to be information recorded about such concepts as the digitization project itself, called the ‘project’; the people who have been involved in the digitization project, referred to as ‘experimenters’; and the make and model of all the hardware involved in the digitization process referred to as the ‘instruments’.

For example, the exact make and model of microscope, microscope camera, and magnification lens were not recorded anywhere. While these pieces of information may not immediately appear relevant, they could still be relevant to future research. For example, a researcher might be curious about something like image quality across multiple microscopes, or maybe identifying watermarks, or digital signatures of particular microscopes to help determine where an unlabeled slide was originally taken. The exact scanning methodology or process is not recorded anywhere, either. Metadata should always be a balance between what is potentially relevant, and what is practical for collection. Therefore, it is possible that in this diatom use case, metadata detailing the scanning process may not be practical for collection. However, hardware does seem like a piece of information worth recording in a metadata form, even if the process or methodology only gets recorded in a paper rather than in the metadata of each scan.

\section{Conclusion and Next Steps}

This paper has reported on metadata research pursued in support of ongoing work to transfer the digital component of the ANS Diatom Herbarium collection to the DataFed platform. Metadata is critical to providing global access to this collection, supporting diatom research activities, and ensuring a robust infrastructure within the ‘Development of a Platform for Accessible Data-Intensive Science and Engineering’ initiative. Background research revealed the value of the Open Microscopy Environment (OME) metadata standard, as well as aspects of the Digital Imaging and Communications in Medicine (DICOM) and Darwin Core (DwC) metadata standard. The baseline metadata alignment examined the complexity of the current ANS Diatom Herbarium metadata structure covering sample metadata, microscope hardware specifications metadata, image acquisition settings metadata, image structure metadata, and scanning metadata and contextualized them by mapping to the following five key types of metadata: descriptive metadata, structural metadata, provenance metadata, technical metadata, and preservation metadata. Additionally, metadata risk analysis revealed a set of concerns with the current state of metadata that cover storage space and potential data loss, metadata loss from image conversion, metadata inconsistency from human error, suppressed metadata maintenance from an insufficient user interface, and missing metadata. 

These efforts have helped our team consider potential solutions, some of which we have already pursed, such as updating the Excel software and improving the image conversion python scripts. These scripts have been prepared, and our next step is to implement them, as we further develop the DataFed mechanism for ANS Herbarium collection on use. 

In moving forward, next steps include addressing administrative aspects of collection sharing and improving the underlying python scripts which convert images to OME-TIFF. Our next steps are guided by a measure that aids diatom researchers both locally and globally, with the overall aim of supporting research advances and knowledge discovery.

\section*{Acknowledgement}
This work is supported by NSF-OAC \#2320600 and Metadata Research Center


\end{document}